\documentclass[showpacs,preprintnumbers]{revtex4-2}
\usepackage{amssymb}
\usepackage{graphicx}
\usepackage{color}

\def\bc{\begin{center}}
\def\ec{\end{center}}

\def\beq{\begin{equation}}
\def\eeq{\end{equation}}

\usepackage{graphicx}
\usepackage{caption}
\usepackage{subcaption}
\usepackage[T1]{fontenc}
\usepackage[utf8]{inputenc}
\usepackage[english]{babel}
\usepackage[hidelinks]{hyperref}
\usepackage{indentfirst}
\usepackage{amssymb}
\usepackage{amsmath}
\usepackage{mathrsfs}
\usepackage{braket}
\usepackage{float}
\usepackage{color}

\begin{document}

\title{Long-living superfluidity of dark excitons in a strip of strained transition metal dichalcogenides double layer}
\author{Gabriel P. Martins$^{1,2,3}$, Oleg L. Berman$^{1,2}$, Godfrey Gumbs$^{2,3}$, and Gabriele Grosso$^{2,4}$}

\affiliation{$^{1}$Physics Department, New York City College
of Technology,          The City University of New York, \\
300 Jay Street,  Brooklyn, NY 11201, USA \\
 $^{2}$The Graduate School and University Center, The
City University of New York, \\
365 Fifth Avenue,  New York, NY 10016, USA\\
 $^{3}$Department of Physics and Astronomy, Hunter College of The City University of New York, 695 Park Avenue, \\
New York, NY 10065, USA\\
$^4$ Photonics Initiative, City University of New York Advanced Science Research Center, 85 St Nicholas Terrace, New York, NY 10031, USA}
\date{\today}

\begin{abstract}

We have proposed the superfluidity of dipolar excitons in  a strip of double-layer transition metal dichalcogenides (TMDCs) heterostructures.  We have shown that strain causes a shift in \textit{k}-space between the minimum of the conduction band and the maximum of the valence band. Therefore, we expect that applying strain to this system can cause dark excitons to be created.  We have numerically calculated the energy spectrum of dark dipolar excitons in strained MoSe$_2$, and we have calculated their binding energies and effective masses. We have shown that the dark dipolar excitons in strained TMDC heterostructures form superfluids, and we have calculated the sound velocity in the energy spectrum of collective excitations, as well as the mean-field critical temperature for superfluidity.  We have shown that two separate superfluid  flows { moving in opposite directions} will appear in the system, one on each edge of the strip, forming the double layer. We have seen that the critical temperature for superfluidity increases with the concentration of dark excitons, as well as with the inter-layer separation. The fact that dark excitons cannot decay by the simple emission of photons, makes it so that the superfluids and condensates formed by them have a much longer lifetime than that formed by bright excitons. {We propose a way to experimentally verify the predicted phenomena.}

\end{abstract}

\maketitle

\section{Introduction}

{ Systems of dipolar, or spatially indirect- excitons, bound states of spatially separated electrons and holes, have been the subject of multiple recent theoretical and experimental works. Such systems are usually formed in coupled quantum wells (CQWs) in semiconductors, or in double layers of two-dimensional  (2D) nanomaterials. Such systems can exhibit Bose-Einstein condensation (BEC) and superfluidity, which can be verified by the existance of persistant electrical currents in each of two layers or through the observation of coherent optical properties \cite{para1,para2,littlewood,para4,para5,para6}. A detailed review on experimental and theoretical works on the superfluidity of dipolar excitons in CQWs can be found in Ref. \cite{para7}. Such superfluidity can happen not only in the BEC regime, but also in the Bardeen-Cooper-Schrieffer (BCS)-BEC crossover regime \cite{para8}.

\medskip
\par

Many theoretical works on the superfluidity of excitons, formed by spatially separated electrons and holes in double-layer graphene  have been accomplished \cite{para9,para10,para11,para12,para125,para13,para14}. Studies of superfluidity of dipolar excitons have also been performed for double layers of transition-metal dichalcogenides (TMDCs) \cite{fabrication0,para16,para17,dipolar1,para19} and double-layer phosphorene  \cite{dipolar2,para21}, which are specially attractive due to the large excitonic binding energies in both cases. A BEC of long-lived dark-spin states of dipolar excitons has been observed in GaAs/AlGaAs semiconductor CQWs \cite{para22}.}

\medskip
\par

In recent years, multiple research groups have been focusing on the study and fabrication of increasingly complex nanodevices \cite{struc1,struc2,struc3,struc4}. Some possible applications of nanostructured devices are their use as sensors, uses for energy storage, energy conversion and many more \cite{sappl1,sappl2,sappl3,sappl4}. The ability to carefully control optoelectronic properties of 2D materials is extremelly important in the fabrication of such devices. Through meticulous engineering of strain fields in 2D materials, one can effectively control some electronic properties of the materials \cite{strainint1,strainint2,strainint3}. 
\medskip
\par

In graphene, for example, the effect due to strain on electrons and holes is similar to that from an external magnetic field \cite{stgraphene1,stgraphene2,stgraphene3}. In graphene, the so-called pseudomagnetic field acts on electrons and holes in a manner that does not depend on the sign of their charges, which leads to the formation of pseudomagneto excitons whose energy dispersion is significantly different from that of magnetoexcitons, when the graphene sheet is subjected to a real magnetic field \cite{oleggraph1,oleggraph2,oleggraph3}. In 2D TMDCs, however, the effect of strain on electrons and holes is much more complex than in graphene.
\medskip
\par

In TMDCs, the effect due to strain is to couple the unperturbed Hamiltonian to five different vector potential fields. These depend on the particular shape and intensity of the strain \cite{guinea1,guinea2,guinea3}. The energy spectrum for electrons in the conduction and valence bands of a strip of TMDC subjected to an arc-shaped strain field has been derived \cite{guinea2}. In this paper, we study excitons in TMDCs under an arc-shaped strain. We will investigate both monolayer excitons, and the more stable double layer - or dipolar -  excitons, when electrons and holes are located on two spatially separated layers of TMDC.

\medskip
\par

In this paper, we will study dipolar excitons in heterostructures consisting of two spatially separated strips of MoSe$_2$ monolayers. Dipolar excitons are formed by electrons and holes in different layers separated by a dielectric, usually hexegonal boron nitride (\textit{h}-BN). Such devices are formed by vertically stacked van der Waals heterostructures, which have been the subject of multiple theoretical and experimental research investigations \cite{fabrication0,fabrication1,fabrication2,fabrication3,fabrication4,fabrication5,fabrication6,fabrication7}. { We treat the excitons in the strip as a quasi-one-dimensional (1D) system. Since excitons are bosonic quasiparticles, they can form Bose-Einstein condensates (BECs). BECs in 1D settings have been discussed in the textbook by Pitaevskii and Stringari \cite{pitbook}.}
\medskip
\par

To create dipolar excitons, a strong electric field perpendicular to the layers is applied in order to force the electron and hole layer separation. The recombination of electrons and holes in such systems is inhibited  by the dielectric barrier between them and the exciton's lifetime is, therefore, much greater than monolayer excitons \cite{lifetime}. {
We show that the strain applied to the TMDC sheets leads to a shift in momentum space between the minimum of the conduction band and the maximum of the valence band. { This means that excitons formed in strained TMDCs are prevented from decaying only through the emission of photons due to momentum conservation.  For that reason, excitons created under conditions such as these are usually referred to as dark excitons \cite{darkexc1,darkexc2,darkexc3,darkexc4,darkexc5}.  } The decay process of dark excitons must involve interaction with phonons. { Due to having a more complex decay process}, dark excitons have a much longer lifetime than bright excitons \cite{darkexc3,darkexc4,darkexc5}, which means that condensates and superfluids of dark excitons also have a longer lifetime. { Dark excitons have been experimentally observed in recent studies involving monolayers with massive anisotropic tilted Dirac-bands \cite{grosso1,grosso2,grosso3}.}

\medskip
\par

It has been predicted that many different systems of dipolar excitons in multilayered heterostructures are expected to exhibit a superfluid phase \cite{littlewood,dipolar1,dipolar2,dipolar3,dipolar4}. This comes from the fact that the energy dispersion of the collective excitations in a weakly interacting Bose gas of dipolar excitons satisfies the Landau's criterion for superfluidity \cite{landau,abrikosov,pitbook}. In this paper, we will demonstrate the  occurrence of superfluidity of  dipolar excitons in strained TMDC strips. { The superfluidity and BEC of dark excitons in double-layer 1T$^\prime$ - MoS$_2$ has been studied in \cite{dipolar4}, where it has been shown that such a system exhibts an angular dependent critical temperature for superfluidity, as well as an angular dependent sound velocity.} {  Superfluidity of excitons in an \textit{h}-BN-separated MoSe$_2$/WSe$_2$ heterostructure has been recently observed \cite{expsec2}.}

\medskip
\par

In this paper, we solved the two-body problem for the electron-hole pair in {a strained strip of double-layer MoSe$_2$}. We have calculated the effective mass and the binding energy for the excitons. {For the weakly interacting Bose gas of dark dipolar excitons, we have obtained} the spectrum of collective excitations and sound velocity. We found the mean-field critical temperature of superfluidity in the system. { The excitons have a doubly degenerate ground state, each located on  opposite edges of the double-layer strip, and with opposite momenta. This leads to the formation of two separate non-interacting superfluid components on the edges of the strip, flowing in opposite directions.   }

\medskip
\par

The remainder of the paper is organized as follows. In Sec.\  \ref{Hamiltonian}, we derive the Hamiltonian for non-interacting pairs of electrons and holes in a TMDC double layer. There, we numerically calculate the energy dispersion of the system as well as obtain the first order energy correction due to the attraction between the electrons and holes. {In Sec. \ref{Hcolsec}, we consider the effects arising from   the exciton-exciton dipole repulsion. We found the energy spectrum of the collective excitations in the weakly interacting Bose gas of dark dipolar excitons. } In Sec. \ref{superfluid}, we demonstrate that such a system will exhibit a superfluid phase and {obtain the mean-field critical temperature for the superfluidity}. { In Sec. \ref{experimental}, we propose an experimental setup that could be used to verify the theoretical predictions made here.} Sec. \ref{conc} concludes our work.

\medskip
\par

\section{Excitons in a Strained strip of double-layer TMDC} 
\label{Hamiltonian}

We consider a structure which consists of electons and holes in two adjacent layers of TMDCs, spatially separated by $N_L$ layers of hexagonal boron nitrate (\textit{h}-BN). We assume that both sheets are subjected to an arc-shaped strain with the same intensity. The considered system is shown schematically in Fig.\  \ref{schem}

\begin{figure}[H]
\includegraphics[width=\textwidth]{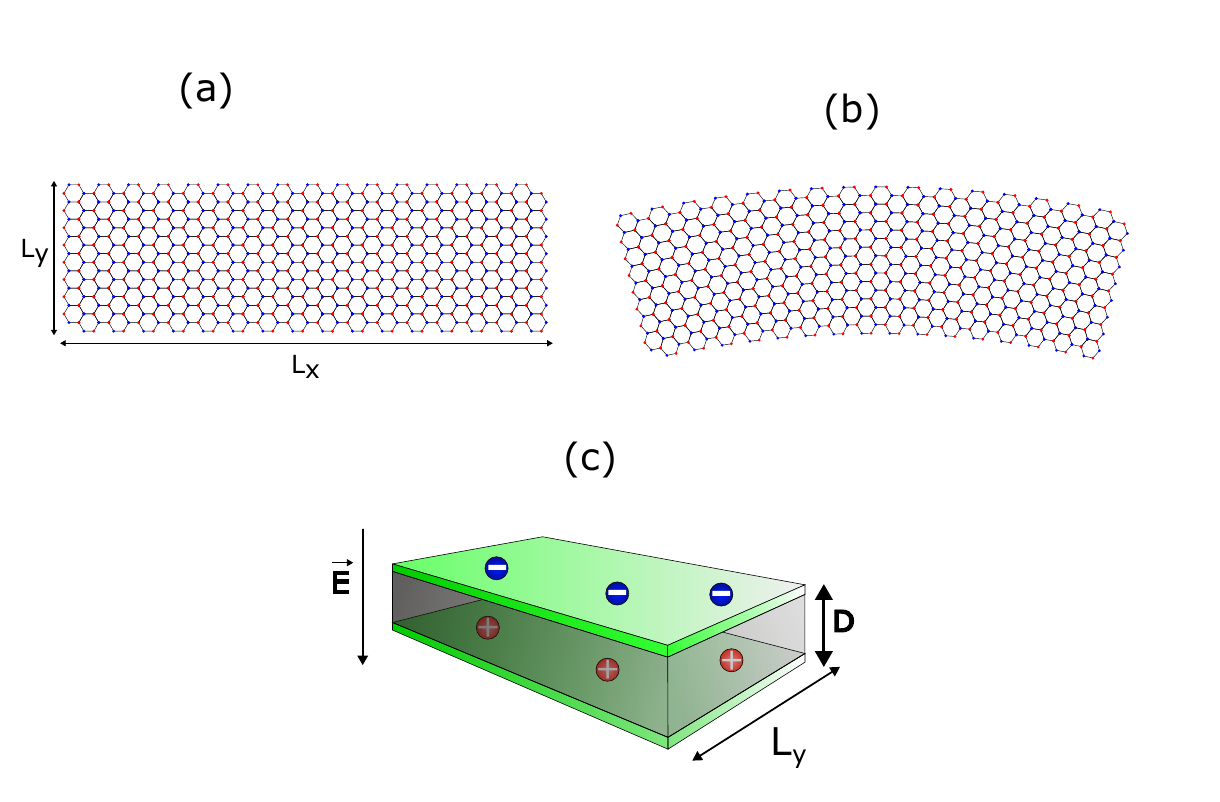}
\caption{(Color online) (a) Schematic illustration of a single strip of unstrained TMDC. The blue sites represent the metal atoms and the red sites contain two chalcogen atoms, in two different sublayers slightly above and below the metal layer. (b) A similar strip as the one in (a), but now subjected to an arc-shaped strain with radius of curvature $R = 3 L_y$. (c) Heterostructures considered throughout this work. Two layers of strained TMDC (depicted in green) are spatially separated by $N_L$ layers of \textit{h}-BN. An external electric field $E$ is applied, to force electrons (blue spheres) and holes (red spheres) to be confined to different layers.}\label{schem}\end{figure}

\medskip
\par

Electrons in a strip of TMDC under an arc-shaped strain {can be described by} a Hamiltonian $\hat{H}$ of the form \cite{guinea2}

\begin{equation}
\hat{H} = \begin{pmatrix}
V_+ (y) - \dfrac{\hbar}{4 m_0}(\alpha+\beta) \partial_y^2 & t_0a_0(q_x+\eta_1\dfrac{y}{a_0 R}) - t_0 a_0 \partial_y \\ t_0a_0(q_x+\eta_1\dfrac{y}{a_0 R}) + t_0 a_0 \partial_y & V_- (y) - \dfrac{\hbar}{4 m_0}(\alpha-\beta) \partial_y^2 
\end{pmatrix}, \label{HFull}\  ,
\end{equation}
where

\begin{equation}
V_\pm = \dfrac{\Delta_0 \pm \Delta}{2} + \dfrac{\hbar^2}{4m_0}\alpha\left(q_x + \eta_2\dfrac{y}{a_0 R}\right)^2 \pm \dfrac{\hbar^2}{4m_0}\beta\left(q_x + \eta_3\dfrac{y}{a_0 R},\right)^2  
\end{equation}
with $m_0$ denoting the free electron mass and $a_0$ is the separation between nearest neighboring atoms in the unstrained TMDC lattice.  Also, $R$ is the radius of curvature of the strain arc, { {$a_0$} is the distance between sublattice sites and the strain-independent parameters $t_0$, $\Delta_0$, $\Delta$, $\alpha$, $\beta$, and $\eta_i$ are  obtained from the Slater-Koster parameters of the unstrained Hamiltonian \cite{guinea2}.}    In the long wavelength limit, one can extract  from Eq.\  (\ref{HFull}) the following effective single-band Hamiltonians for electrons in the conduction ($H_+$) and valence ($H_-$) bands \cite{guinea2}

\begin{eqnarray}
H_+ = E_+ +\dfrac{\hbar^2}{4m_0}\left[(\alpha+\beta+\gamma)q_y^2+\dfrac{w_1^+}{(a_0R)^2}(y-w_2^+a_0R q_x)^2 + w_3^+ q_x^2\right] \label{Hcond}\\
H_- = E_- +\dfrac{\hbar^2}{4m_0}\left[(\alpha-\beta-\gamma)q_y^2+\dfrac{w_1^-}{(a_0R)^2}(y-\tau w_2^-a_0R q_x)^2 + w_3^- q_x^2\right],
 \label{Hval}
\end{eqnarray}
where $E_\pm = \dfrac{\Delta_0 \pm \Delta}{2} \pm \eta_1\dfrac{(t_0a_0)^2}{\Delta a_0 R}$; $\tau = \pm 1$ is the valley index; and $w_i^\pm$ and $\gamma$  are strain-independent variables. { In this paper, we consider a strip of strained double-layer  MoSe$_2$, and we used the values of all strain-independent parameters from Ref. \cite{guinea2}. Namely, we assumed $t_0 = 2.34$ eV, $a_0 = 2.19$ \r{A}, $\Delta_0 = -0.11$ eV, $\Delta = 1.82$ eV, $\alpha = -0.01$, $\beta = -1.54$, $\eta_1 = 0.002$, $\eta_2 = -56.551$, $\eta_3 = 1.635$, $\omega_1^+ = -32.2$, $\omega_1^- = -24.0$, $\omega_2^+ = 0.06$, $\omega_2^- = 0.12$, $\omega_3^+ = 4.05$, and $\omega_3^- = -3.57$. } The energy spectrum of the electrons in the conduction and valence band of MoSe$_2$ under an arc-shaped strain was numerically calculated in Ref. \cite{guinea2}. We show this result in Fig. \ref{edispcondval} (a).

\begin{figure}[H]
\begin{center}
\begin{subfigure}[b]{0.40\textwidth}
\includegraphics[width = \textwidth]{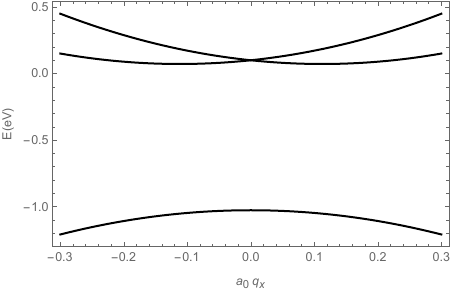}
\caption{}
\end{subfigure}
\begin{subfigure}[b]{0.40\textwidth}
\includegraphics[width = \textwidth]{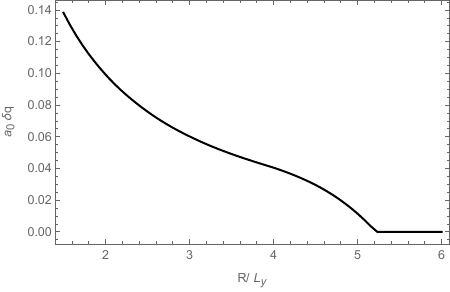}
\caption{}
\end{subfigure}
\end{center}
\caption{(a) Energy dispersion of the conduction and valence bands around the $K$ and $K^\prime$ points for electrons in a strip of strained MoSe$_2$. In here, we considered $L_y = 50$ $a_0$, and $R = 1.8L_y$. (b) Difference $\delta q$ between the values of $q_x$ for the maximum of the valence band (which is always at $q_x = 0$), and the minimum of the conduction band, as a function of the radius of curvature of the strain $R$, for a strip of MoSe$_2$ where $L_y$ = 50 $a_0$.}\label{edispcondval}
\end{figure}

\medskip
\par

It is clear from Fig. \ref{edispcondval}(a) that the maximum of the valence band and the minima of the conduction band do not occur at the same value of the wave vector along the $q_x$ direction. This means that the decay process of electrons at the minimum of the conduction band to the maximum of the valence band cannot take place through the emission of a photon alone, since it would violate momentum conservation. Such a decay must be mediated by phonons as well. Excitons created between electrons in the conduction band and holes in the valence band are, therefore, much more resilient to decay. In situations like these, when momentum conservation prohibits the decay of excitons by the simple emission of photons, excitons in the system are called dark excitons. 

\medskip
\par

In Fig. \ref{edispcondval}(b), we see how the difference in momentum $q_x$ between the minimum of the conduction band and the maximum of the valence band depend on the strain-caused curvature $R$ of the strip.  From Fig. \ref{edispcondval}(b), we notice that there exists a minimum amount of strain necessary in order for excitons in the system to be dark. In other words, if the strain is small, such that $R>R_c$, the minimum of the conduction band and the maximum of the valence band both exist when $q_x = 0$, and the excitons created in such a system are bright, and can decay through the emission of photons, without phonom interactions. For a strip of MoSe$_2$ with $L_y = 50$ $a_0$, this phase transition between dark and bright excitons happens around $R_c \approx 5.2$ $L_y$.
\medskip
\par

The Hamiltonian for holes $H_h$ in the valence band can be extracted from $H_-$ by applying an overall change of sign in the Hamiltonian, together with a change in the sign of the momenta $q_i$ {($i = x,y$)}:

\begin{equation}
H_h = -E_- + \dfrac{\hbar^2}{4m_0}\left[(\beta+\gamma-\alpha)q_y^2 - \dfrac{w_1^-}{(a_0R)^2}(y+\tau w_2^- a_0Rq_x)^2-w_3^- q_x^2 \right]. \label{Hhole}
\end{equation}
Since both $w_1^-$ and $w_3^-$ are negative constants \cite{guinea2}, the Hamiltonian for holes shown in Eq. \ref{Hhole} is that of a particle in a harmonic potential. The energy dispersion for holes in strained TMDCs is given by $E_h(n,q_x) = -\left(E_--\dfrac{\hbar^2}{4m_0}w_3^-q_x^2\right) + \hbar\omega\left(n +\dfrac{1}{2}\right)$. The lowest energy state for holes in the valence band happens at $q_x = 0$, which is a point in which the Hamiltonian in Eq. (\ref{Hhole}) is the same for both intra-valley and inter-valley electrons and holes.

\medskip
\par

The Hamiltonian $\mathcal{H}_0$ for non-interacting excitons will be, therefore, given by $\mathcal{H}_0 = H_e + H_h$, where we relabeled the Hamiltonian for conduction electrons as $H_e$ for clarity. This Hamiltonian can be 
written as

\begin{eqnarray}
\mathcal{H}_0 &=& H_e + H_h 
\nonumber \\
&=& \Delta E +\dfrac{\hbar^2}{4m_e} {q_y}_e^2 + \dfrac{\hbar^2}{4m_h} {q_y}_h^2 
\nonumber \\
&+&\dfrac{\hbar^2}{4m_0}\left[\dfrac{w_1^+}{(a_0R)^2}(y_e-w_2^+a_0R{q_x}_e)^2-\dfrac{w_1^-}{(a_0R)^2}(y_h+w_2^-a_0R{q_x}_h)^2 + w_3^+{q_x}_e^2 -w_3^-{q_x}_h^2 \right], \label{h0ruim}
\end{eqnarray}
where $m_e = \dfrac{m_0}{\alpha+ \beta + \gamma}$ {is the effective mass of the electron, $m_h = \dfrac{m_0}{\beta + \gamma - \alpha}$ is the effective mass of the hole}, and $\Delta E = E_+-E_-$. 

{We can separate the center-of-mass motion by performing the following change of variables}
\begin{eqnarray}
\mathbf{Q} = \mathbf{q}_e + \mathbf{q}_h  &;& \mathbf{R} = \dfrac{m_e\mathbf{r}_e+m_h\mathbf{r}_h}{m_e+m_h} \nonumber \\
\mathbf{q} = \dfrac{m_h\mathbf{q}_e - m_e\mathbf{q}_h}{m_e+m_h} &;& \mathbf{r} = \mathbf{r}_e - \mathbf{r}_h,
\end{eqnarray} 
{ where $(\mathbf{q}_e,\mathbf{r}_e)$, and $(\mathbf{q}_h,\mathbf{r}_h)$ are the canonical coordinates of the electron and hole, respectively; and $(\mathbf{Q},\mathbf{R})$, and $(\mathbf{q},\mathbf{r})$ are the canonical coordinates of the center of mass and of the relative motion for the electron-hole pair, respectively.}

With these new coordinates, Eq. (\ref{h0ruim}) becomes
\begin{equation}
\mathcal{H}_0 = \Delta E +\dfrac{\hbar^2}{4M} Q_y^2 + \dfrac{\hbar^2}{4\mu} q_y^2 + V(Y,y,Q_x,q_x), \label{Hnonint}
\end{equation}
where {$M = m_e+m_h$ is the exciton's total mass, $\mu = \dfrac{m_e m_h}{m_e+m_h}$ is the exciton's reduced mass}, and $V$ is given by
\begin{eqnarray}
V(Y,y,Q_x,q_x) &=& \dfrac{\hbar^2}{4 m _0}\left\lbrace C_1  Y^2 + C_2Yy+C_3 y^2 + C_4YQ_x \right. \nonumber \\
& & \left.+ C_5Yq_x + C_6 yQ_x+ C_7 yq_x + C_8q_x^2 + C_9 Q_xq_x +C_{10}Q_x^2 \right\rbrace, \label{potential}
\end{eqnarray}
where
\begin{eqnarray}
C_1 = \dfrac{w_1^+-w_1^-}{(a_0R)^2} ; & C_2 = - 2\dfrac{w_1^+ m_h+w_1^- m_e}{M(a_0R)^2} ; & C_3 = \dfrac{w_1^+m_h^2-w_1^-m_e^2}{(Ma_0R)^2} \nonumber \\
C_4 = -\dfrac{\beta^+ m_e+\beta^- m_h}{M}; & C_5 = \beta^--\beta^+ ; & C_6 = \dfrac{\mu}{M}(\beta^--\beta^+) \nonumber \\
C_7 = -\dfrac{\beta^+ m_h + \beta^- m_e}{M}; & C_8 = \alpha^+ -\alpha^- ;& C_9 = 2\dfrac{\alpha^+ m_e + \alpha^- m_h}{M} \nonumber \\
&C_{10} = \dfrac{\alpha^+ m_e^2 - \alpha^- m_h^2}{M^2}, &
\end{eqnarray}
in which $\alpha^\pm = w_1^\pm {w_2^\pm}^2 +w_3^\pm$, and $\beta^\pm = 2w_1^\pm w_2^\pm/(a_0R)$. Since $\mathcal{H}_0$ is independent of both $x$ and $X$, $q_x$ and $Q_x$ are both integrals of motion for the Hamiltonian of non-interacting electrons and holes. The energy dispersion for such system is shown in Fig. \ref{nonint}.

\begin{figure} [H]
\begin{center}
\includegraphics[width=0.5\textwidth]{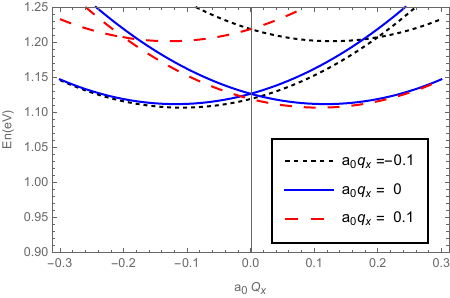}
\end{center}
\caption{Energy dispersion for non-interacting excitons in strained MoSe$_2$. The first two energy levels are being shown for each value of $q_x$. }\label{nonint}
\end{figure}

{ In Fig. \ref{nonint}, we again see the evidence of dark excitons which should be created in this system. It is clear from the figure that excitons in the ground state have non-zero total linear momentum $Q_x$. In order for such an exciton to be created, the absorption of a photon by electrons in the valence band is not enough, since that would violate momentum conservation. This means that the creation and annihilation of excitons in TMDCs under arc-shaped strain must involve interactions with phonons as well as with photons. 

\medskip
\par

{
The excitons have a doubly degenerate ground state, { which appears at specific values of the quantum numbers $q_x$ and $Q_x$. The ground state for the excitons corresponds to $q_x = \pm q_{min}$ and $Q_x = \pm Q_{min}$}. For the values considered here  for the physical parameters (a width $L_y = 50$ $ a_0$, and a radius of curvature  $R = 1.8$ $ L_y$), we have $q_{min} a_0 \approx 0.056$ and $Q_{min}a_0 \approx 0.1122$.  In Fig.\  \ref{Veff}, we show the effective potential given by Eq.\ (\ref{potential}) experienced by excitons in each of the ground states as a function of the position of the exciton's center of mass alongside the width of the strip $Y$.

\begin{figure}[H]
\begin{center}
\includegraphics[width = 0.35\textwidth]{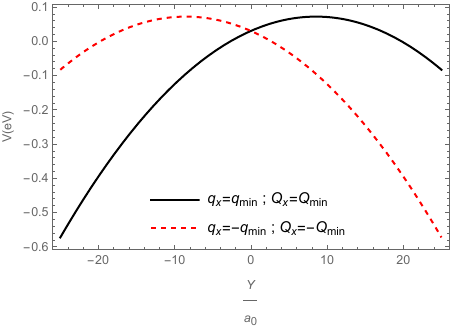}
\end{center}
\caption{(Color online) Effective potential given by Eq. (\ref{potential}) along the width of the strip for excitons at the $q_x = q_{min}$ and $Q_x = Q_{min}$ ground state (solid black line, and $q_x = -q_{min}$ and $Q_x = -Q_{min}$ ground state (dashed red line). In both cases, we consider $L_y$ = 50 $a_0$ and $R = 1.8 $ $L_y$. \label{Veff}}
\end{figure}
{
As it can be seen from Fig. \ref{Veff}, excitons in different ground states experience effective potential whose minimum lies in opposite ends of the double-layer strip. This means that the excitons will also be localized around opposite ends of the strip.
}}

}

The full Hamiltonian $\mathcal{H}$ for interacting electrons and holes is given by
\begin{equation}
\mathcal{H} = \mathcal{H}_0 +V (r_{eh}), \label{Hamnice}
\end{equation}
where $V (r_{eh})$ is the interaction potential for electrons and holes, located at a distance $r_{eh}$ from one another. In the monolayer case, $V$ is the Rytova-Keldysh potential $V_{RK}$ given by \cite{keldysh1,keldysh2,rubio}
\begin{equation}
V_{RK} (r_{eh}) = - \dfrac{\pi k e^2}{\epsilon_d \rho_0}\left[ H_0\left(\dfrac{r_{eh}}{\rho_0}\right)-Y_0\left(\dfrac{r_{eh}}{\rho_0}\right)\right], \label{Vkel}
\end{equation}
where $H_0$ and $Y_0$ are Struve and Bessel functions of the second kind, respectively; $\epsilon_d$ is the relative permitivity of the dielectric medium, and $\rho_0$ is the screening length. In the limit of large electron-hole separations ($r_{eh}\gg \rho_0$), the potential given in Eq. (\ref{Vkel}) behaves like the Coulomb potential $V_C =-\dfrac{ke^2}{\epsilon_d r_{eh}}$. For double layers of TMDCs, the interaction can be approximated by the Coulomb potential with minimal loss of information \cite{rubio}. The energy correction due to the interaction potential $V$ in the eigenstates of $\mathcal{H}_0$ can be calculated using first order perturbation theory \cite{landaubook}.  The energy up to first order corrections will be given by
\begin{equation}
E^1(Q_x,q_x) = E_0(Q_x,q_x) + \delta E ^\prime_1(Q_x,q_x),
\end{equation} 
where $E_0(Q_x,q_x)$ are the energy eigenvalues of the non-interacting electrons and holes obtained from Eq. (\ref{Hnonint}) (the ones shown in Fig. \ref{nonint}). {Since the energy corrections $\delta E ^\prime_1(Q_x,q_x)$ are always negative, we will instead treat the positive values defined by $\delta E _1(Q_x,q_x) = - \delta E ^\prime_1(Q_x,q_x)$, for clarity. } The first order corrections $\delta E_1(Q_x,q_x)$ are given by
\begin{equation}
\delta E_1(Q_x,q_x) = - \bra{\Psi^0_{Q_x,q_x}}V\ket{\Psi^0_{Q_x,q_x}},
\end{equation}
where $\ket{\Psi^0_{Q_x,q_x}}$ are the eigenvectors of $\mathcal{H}_0$ obeying $\mathcal{H}_0\ket{\Psi^0_{Q_x,q_x}} = E_0(Q_x,q_x)\ket{\Psi^0_{Q_x,q_x}}$. The binding energy $\varepsilon_b$ is the first order correction $\delta E_1$ to the energy of the ground state of $\mathcal{H}_0$, which, for $R<R_c$, does not happen at $Q_x = q_x = 0$. 

Throughout the remainder of this paper, we consider heterostructures consisting of two layers of MoSe$_2$, separated by $N_L$ layers of hexagonal boron nitride ($h$-BN). In this case, the distance between the MoSe$_2$ layers will be $D=N_L * 0.33$ nm, and the relative permitivity of $h$-BN is $\epsilon_d = 4.89$. We consider the width of the strip to be $L_y = 50$ $a_0$. Unless specifically written otherwise, we consider the inter-layer separation to consist of $N_L = 10$ layers of $h$-BN ($D = 3.33$ nm), and the radius of curvature of strain to be $R = 1.8$ $L_y$.

\medskip
\par 
In Fig. \ref{CoulKel}, we compare the binding energies for the excitons as a function of the inter-layer separation $D$ when we treat the interaction between electrons and holes by either the Coulomb and Rytova-Keldysh potential. In Fig.\  \ref{encor}, we show how the first order energy corrections $\delta E_1$ change from intra-valley excitons (when electrons and holes are in the same valley) to inter-valley excitons (when electrons and holes are in different valleys).

\begin{figure}[h]
\includegraphics[width = 0.35\textwidth]{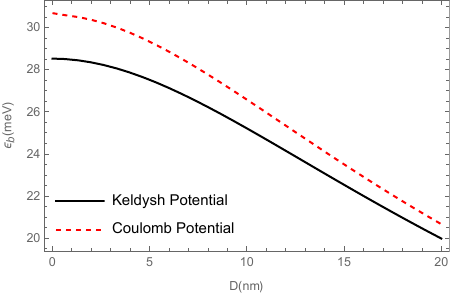}
\caption{(Color online) Excitonic binding energy as a function of the inter-layer separation $D$ using the Rutova-Keldysh (solid black curve) and the Coulomb (dashed red curve) potentials. The plot for binding energies up to first order corrections is identical for both inter-valley excitons and intra-valley excitons.}
\label{CoulKel}
\end{figure}

We can see from Fig.\  \ref{CoulKel} that the binding energies found when considering the interactions between electrons and holes to be governed by the Coulomb and Rytova-Keldysh potentials are quite similar and approach the same asymptotic value at large inter-layer separation $D$. We see that the difference between the binding energies is greatest for monolayer excitons (when $D=0$), which is exactly when the two potentials differ the most.

\begin{figure}[H]
\begin{center}
\begin{subfigure}[b]{0.30\textwidth}
\includegraphics[width=\textwidth]{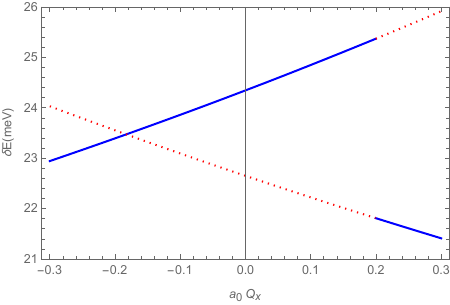} \\
\includegraphics[width=\textwidth]{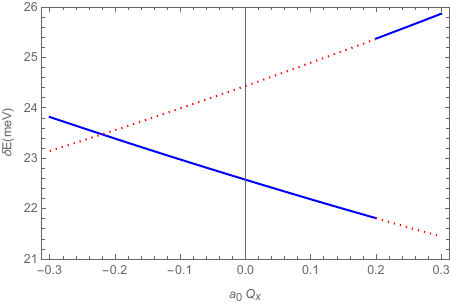}
\caption{$a_0q_x = -0.1$}
\end{subfigure}
\begin{subfigure}[b]{0.30\textwidth}
\includegraphics[width=\textwidth]{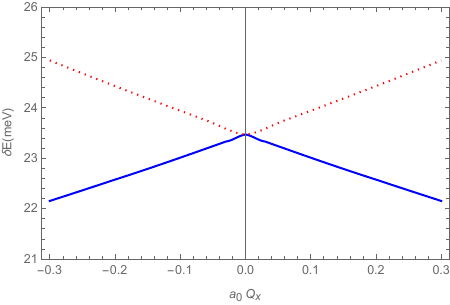} \\
\includegraphics[width=\textwidth]{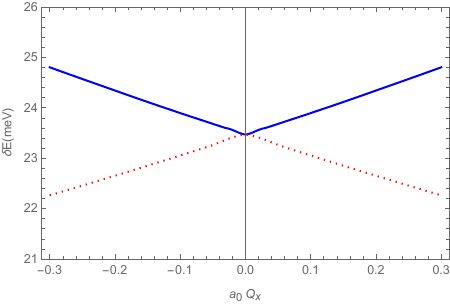}\caption{$a_0q_x = 0$}
\end{subfigure}
\begin{subfigure}[b]{0.30\textwidth}
\includegraphics[width=\textwidth]{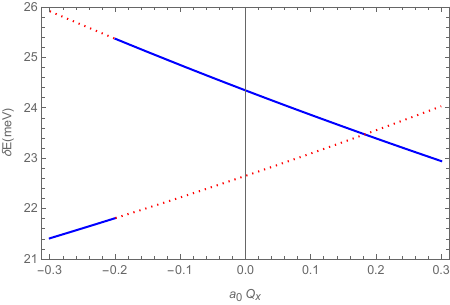} \\
\includegraphics[width=\textwidth]{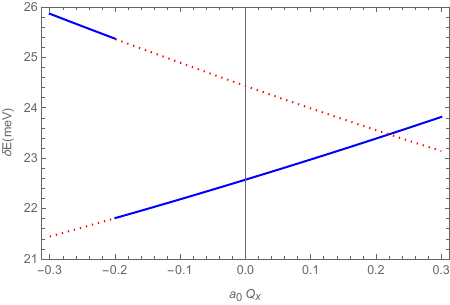}
\caption{$a_0q_x = 0.1$}
\end{subfigure}
\end{center}
\caption{(Color online) Corrections to the energy of non-interacting electrons and holes due to the Rytova-Keldysh potential using first order perturbation theory for various values of $a_0q_x$, calculated for an inter-layer separation of $D = 3.33$ nm (corresponding to $N_L=10$ layers of \textit{h}-BN between the two MoSe$_2$ layers). The solid blue line corresponds to the correction to the energy of the first Landau level and the dashed red line to the correction to the energy of the second Landau level. On the top, we have the corrections to the energy of Intra-valley excitons (when the electron and hole are on the same valley), and on the bottom to the Inter-valley excitons (when electron and hole are on different valleys). \label{encor}}
\end{figure}

\medskip
\par

From Fig.\  \ref{encor}, we also observe that the difference in the first order corrections to the energy of interacting electrons and holes to inter-valley and intra-valley excitons is, at most, of order 3.0 meV. Given that the energies of non-interacting electrons and holes are of the order of 1.0 eV, this means that the difference in the energy dispersion of intra-valley and inter-valley is, at most, of 0.3 $\%$. The energy dispersion can, therefore, be considered the same for both types of excitons in strained TMDC with minimal error.

\begin{figure}[H]
\begin{center}
\begin{subfigure}[b]{0.30\textwidth}
\includegraphics[width=\textwidth]{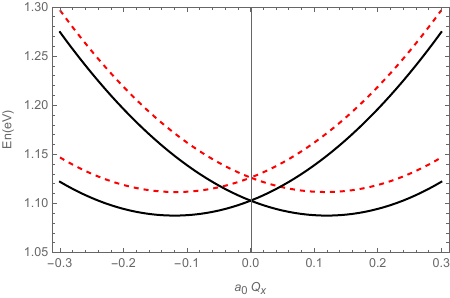}
\caption{$a_0q_x = 0$}
\end{subfigure}
\begin{subfigure}[b]{0.30\textwidth}
\includegraphics[width=\textwidth]{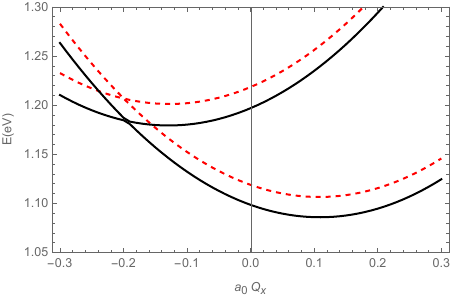
}
\caption{$a_0q_x = 
0.1$}
\end{subfigure}
\begin{subfigure}[b]{0.30\textwidth}
\includegraphics[width=\textwidth]{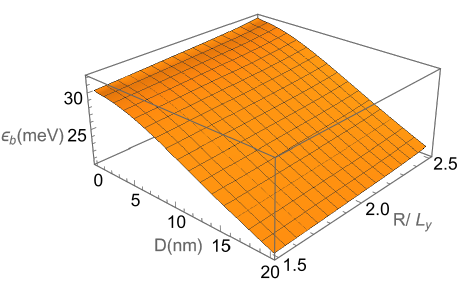}
\caption{}
\end{subfigure}
\end{center}
\caption{(Color online)(a,b) Energy dispersion for non-interacting (dashed red line) and interacting (solid black line) electrons and holes for two values of $a_0q_x$; (c) binding energy of dipolar excitons as a function of the inter-layer distance $D$, and of the radius of curvature of strain $R$. \label{ex_spec}}
\end{figure}

\medskip
\par 
{ From Fig.\ \ref{ex_spec}, it is evident that the addition of the electron-hole attraction to the Hamiltonian does not substantially influence the qualitative behavior of the energy dispersion. It does brings the excitonic energy down by a few meV, as it should be expected, but, at least at first order, it does not break the degeneracies of the energy spectrum  as evidenced in the crossings of the lines in Figs. \ref{ex_spec}(a) and (b). We can also see from Fig.\  \ref{ex_spec}(c) that the excitonic binding energy decreases sharply with the distance between the two TMDC layers $D$, and mildly with the intensity of strain (which is inversely proportional to $R$).}

\section{Energy Spectrum of Collective excitations} 
\label{Hcolsec}

In the low temperature limit, the energy levels which are accessible to the excitons will be the ones near each of the doubly degenerate ground states, in which $q_x = \pm q_{min}$, and $Q_x = \pm Q_{min} +  Q$, with $Q$ small ($a_0Q\ll 1$). In this case, the energy of the excitons will be determined by

\begin{equation}
\varepsilon_0( Q) = \varepsilon_0 + \dfrac{\hbar^2 Q^2}{2M}, \label{en_mass}
\end{equation}  
where $\varepsilon_0$ is the energy of the excitonic ground state, and $M$ is the exciton's effective mass. The energy $\varepsilon_0$ in Eq. (\ref{en_mass}) represents the energy required to create an exciton in the strained TMDC. It is a constant energy present in the spectrum that can be, for simplicity, relabeled to zero ($\varepsilon_0 \rightarrow 0$ from this point onward). We have numerically calculated the effective mass of dipolar excitons for various values of $R$ and $D$ and we noted that they remained effectively unchanged at $M = 1.07$ $ m_0$.


{ Excitons with $q_x \approx q_{min}$ and $Q_x \approx Q_{min}$, and excitons with $q_x \approx - q_{min}$ and $Q_x \approx - Q_{min}$ are located on opposite edges of the TMDC strip. {This means that the distance between excitons at different ground states is much larger than the distance between excitons at the same ground state.  We can, therefore, consider only the interaction between excitons located in $k$-space in the vicinity of the same ground state. We can, therefore, treat the two separate ground states effectively as separate, non-interacting systems. Both with the same energy dispersion relation, but located at opposite edges of the TMDC strip{, and moving in opposite directions.}}}

{Excitons are not elementary particles, but are composite bosons \cite{abrikosov}. They differ from regular bosons only due to the effects of exchange interactions between the electrons and holes. At the limit of low densities and large inter-layer separations $D$, however, these interactions can be neglected \cite{dipolar2}. In this limit, we can treat the system of dipolar excitons, {located at each of the edges of the strained double-layer TMDC strip,}  as a weekly interacting Bose gas, via the effective Hamiltonian \cite{dipolar1,dipolar2,dipolar3,dipolar4,abrikosov}
}
\begin{equation}
\hat{H}_i = \sum_{Q} \varepsilon_0(Q)\hat{a}^\dagger_{Q_i}\hat{a}_{Q_i} + \dfrac{g_i}{L_x}\sum_{Q_1,Q_2,Q_3,i}\hat{a}^\dagger_{Q_{1_i}}\hat{a}^\dagger_{Q_{2_i}}\hat{a}_{Q_{3_i}}\hat{a}_{{(Q_1+Q_2-Q_3)}_i}, \label{Ham_2q}
\end{equation}
{where $i=$ $L,R$ represent the left and right {edges of the strip,} $\hat{a}^\dagger_{Q_i}$ and $\hat{a}_{Q_i}$ are the creation and annihilation operators for excitons with momentum $Q$ at the $i$ edge, $L_x$ is the length of the strip in the $x$ direction, and $g_i$ is a coupling constant due to the repulsive interaction between the excitons.  
 
At low temperatures, almost all of the excitons will be in the ground state. In this case, we can apply Bogoliubov's approximation \cite{abrikosov,fetter}. The chemical potential $\mu$ can be found by minimizing $\hat{H}_{0_i}-\mu\hat{N_i}$ with respect to the one-dimensional concentration of dipolar excitons $n_i = N_i/L_x$, where $N_i$ is the number of excitons at the $i$ edge, and $\hat{H}_{0_i}$ is the Hamiltonian in Eq. (\ref{Ham_2q}) describing particles in the condensate (when $Q = 0$), and the number operator $\hat{N}_i$ is given by
\begin{equation}
\hat{N}_i = \sum_{Q,j} \hat{a}^\dagger_{Q_i}\hat{a}_{Q_i}.
\end{equation}
From this point onward, for simplicity of notation, we will be omitting the index $i$ that differentiate the left and right edges from the equations, as the equations for left and right edges are identical to one another. It is important to note, however, that, in principle, it is possible to control the exciton concentration in each edge separately as there is no physical constraint that imposes $n_L = n_R$.  }
We can solve for $\mu$ by replacing the operator $\hat{N}$ by $N = nL_x$, which leads to

\begin{equation}
\mu =  gn. \label{eqmu}
\end{equation}
We can find the value of $g$ by taking an approach similar to the one in Refs. \cite{dipolar1,dipolar2,dipolar3,dipolar4,abrikosov}. In the dilute regime, the distance $r$ between two excitons cannot be less than the classical turning point, in other words, the energy of two excitons cannot exceed twice the chemical potential

\begin{equation}
U(r_0) \approx 2\mu, \label{tpoint}
\end{equation}
where $U(r_0)$ is the potential energy at a distance $r_0$, and $r_0$ is the { distance between two dipolar excitons at their classical turning point.}

\medskip
\par 
The potential energy  $U(r)$  between two excitons will be given by,

\begin{equation}
U(r) = 2V(r) - 2V\left(r\sqrt{1+\dfrac{D^2}{r^2}}\right), \label{Ufull}
\end{equation}
where $V(r)$ is the potential energy between two electrons or two holes in the same TMDC layer. { In the dilute regime, we assume that} $r\gg D, \rho_0$, and, in this case, we can consider the interaction potential $V(r)$ to be the Coulomb potential $V = \dfrac{ke^2}{\epsilon_d r}$. By expanding Eq. (\ref{Ufull}) and keeping only the lowest order term in $D/r$, we find \cite{dipolar2}
\begin{equation}
U(r) = \dfrac{ke^2D^2}{\epsilon_dr^3}. \label{Ubom}
\end{equation}

The value of $g$ can be found by following  an approach similar to the one in Refs.\   \cite{dipolar1,dipolar2,dipolar3}
\begin{equation}
g = \int_{r_0}^\infty U(r)dr,
\end{equation}
which leads to
\begin{equation}
g = \dfrac{ke^2D^2}{2\epsilon_dr_0^2}.\label{eqg}
\end{equation}
Combining Eqs. (\ref{eqmu}), (\ref{tpoint}), (\ref{Ubom}), and (\ref{eqg}), we find
\begin{eqnarray}
r_0 &=& \dfrac{1}{n}; \\
g &=& \dfrac{ke^2 D^2n^2}{2\epsilon_d}.
\end{eqnarray}

We can now go to the framework of Bogoliubov's approximation, ideal to the study of weakly interacting Bose gases. Instead of using the Hamiltonian in Eq. (\ref{Ham_2q}), we will then consider the Hamiltonian for collective excitations $\hat{H}_{col}$, given by
\begin{equation}
\hat{H}_{col} = \sum_{Q\neq 0,j} \varepsilon(Q) \hat{\alpha}_{Q,j}^\dagger\hat{\alpha}_{Q,j}, \label{Ham_col}
\end{equation}
where $\hat{\alpha}_{Q,j}^\dagger$ and $\hat{\alpha}_{Q,j}$ are the creation and annihilation operators for the Bose quasiparticles with energy dispersion given by
\begin{equation}
\varepsilon(Q) = \left[\left(\varepsilon_0(Q)+\mu\right)^2-\mu^2\right]^{1/2}. \label{disp_col}
\end{equation}
In the first order in $Q$, Eq. (\ref{disp_col}) becomes
\begin{equation}
\varepsilon(Q) = \hbar \sqrt{\dfrac{gn}{2M}}Q. \label{disp_col2}
\end{equation}
The energy dispersion in Eq. (\ref{disp_col2}) is linear. { This means that we can rewrite Eq. (\ref{disp_col2}) as $\varepsilon(Q) = \hbar c_s Q$, with the sound-velocity  $c_s = \sqrt{\dfrac{gn}{2M}}$.}

\begin{figure}[H]
\begin{center}
\begin{subfigure}[b]{0.40\textwidth}
\includegraphics[width=\textwidth]{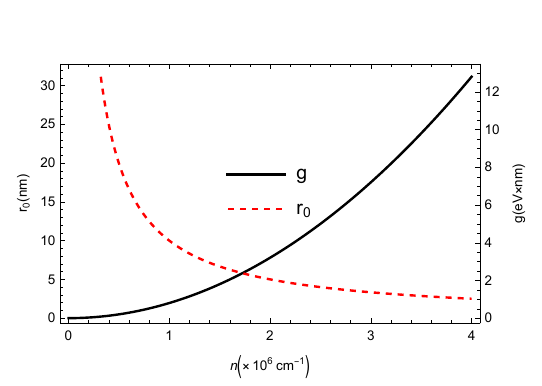}
\caption{}
\end{subfigure}
\begin{subfigure}[b]{0.40\textwidth}
\includegraphics[width=\textwidth]{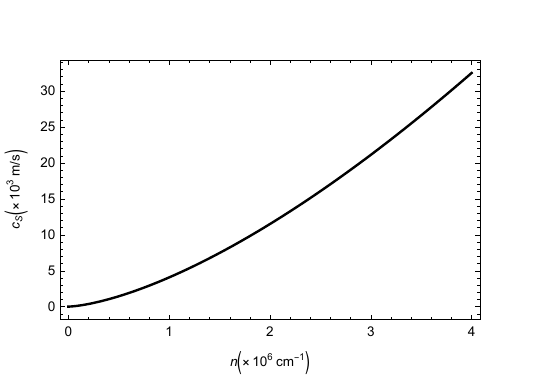}
\caption{}
\end{subfigure}

\end{center}
\caption{(Color online)(a) Values of $g$ (solid black line) and $r_0$ (dashed red line) as a function of the exciton concentration $n$; (b) sound velocity $c_S$ as a function of the exciton concentration $n$. Both graphs were calculated for a system with $N_L = 7$ layers of h-BN between  two strained TMDC layers. \label{rgc}}
\end{figure}

{ In Fig. \ref{rgc}(a), we see the numerically calculated values for the { distance $r_0$ between two dipolar excitons at their classical turning point},  and the coupling strength $g$. In Fig. \ref{rgc}(b), we present the numerically calculated sound velocity $c_S$, both as a function of the one-dimensional exciton concentration $n$. It is clear that both the sound velocity and coupling strength monotonically increase with the exciton concentration, while  {the value of $r_0$}  monotonically decreases with it.}

\section{Superfluidity of Dipolar Excitons} 
\label{superfluid}

As we have seen from Eq.\ (\ref{disp_col2}), the spectrum of dipolar excitons in strained TMDC is sound-like. This means that such a system satisfies the Landau criterion for superfluidity \cite{pitbook,landau}, with the critical velocity for superfluidity being the sound velocity $c_S$. The density $\rho_S$ of the superfluid component at temperature $T$ satisfies $\rho_S(T) = \rho - \rho_n(T)$, where $\rho = Mn$ is the total density of excitons in the TMDC and $\rho_n(T)$ is the density of the normal component. We will calculate the normal density $\rho_n(T)$ following the usual approach in Refs.\  \cite{dipolar1,dipolar2,dipolar3}.

\medskip
\par 
We now assume that the exciton gas is moving through the sample with relative velocity $u$. This means that the superfluid components is moving with velocity $u$. For nonzero temperatures, dissipating quasiparticles will appear in the system. At low temperatures, the number of quasiparticles will be small and they can be treated as an ideal Bose gas. Let us calculate the mass current $J$ in the frame of reference of the superfluid component. The mass current is given by

\medskip
\par 

\begin{equation}
J = s\hbar^2 \int \frac{dQ}{2\pi\hbar} Qf[\varepsilon(Q)-\hbar Q u], \label{J1}
\end{equation}
where $f(\varepsilon) = [\exp(\varepsilon/k_BT)-1]^{-1}$ is the Bose-Einstein distribution function for the dissipative quasiparticles, { and $s = 16$ is the spin and valley degeneracy factor}. We can expand the integral in Eq. (\ref{J1}) in terms of $\hbar Qu/k_BT$ and keep only the first order term and obtain  

\begin{equation}
J = -\hbar^3 s u \int	\frac{dQ}{2\pi\hbar} \  Q^2\dfrac{\partial f[\varepsilon(Q)]}{\partial \varepsilon(Q)}. \label{J2}
\end{equation}
The normal density $\rho_n$ is defined by \cite{pitbook}
\begin{equation}
J = \rho_n u. \label{J3}
\end{equation}
Combining Eqs. (\ref{J2}) and (\ref{J3}), we have the following equation for the density of the normal phase
\begin{equation}
\rho_n = -\hbar^3 s \int \dfrac{dQ}{2\pi\hbar} Q^2\dfrac{\partial f[\varepsilon(Q)]}{\partial \varepsilon(Q)}. \label{Eqrhon}
\end{equation}
By solving Eq. (\ref{Eqrhon}), we find
\begin{equation}
\rho_n(T) = \dfrac{I s(k_BT)^2}{\pi\hbar c_s^3},
\end{equation}
where $I = \int_0^\infty \frac{x^2e^x}{(e^x-1)^2} dx\approx 3.29$. The superfluid concentration $n_s = \rho_s/M$ is, therefore, given by
\begin{equation}
n_s = n - \dfrac{I s (k_BT)^2}{\pi\hbar M c_s^3}.
\end{equation}

The mean-field critical temperature for superfluidity $T_c$ is the temperature at which all of the excitons are in the normal component $\rho_n(T_c) = \rho = Mn$, which leads to
\begin{equation}
T_c = \sqrt{\dfrac{\pi\hbar c_s^3 Mn}{I s k_B^2}}.
\end{equation}
\begin{figure}[h]
\begin{center}
\begin{subfigure}[b]{0.40\textwidth}
\includegraphics[width=\textwidth]{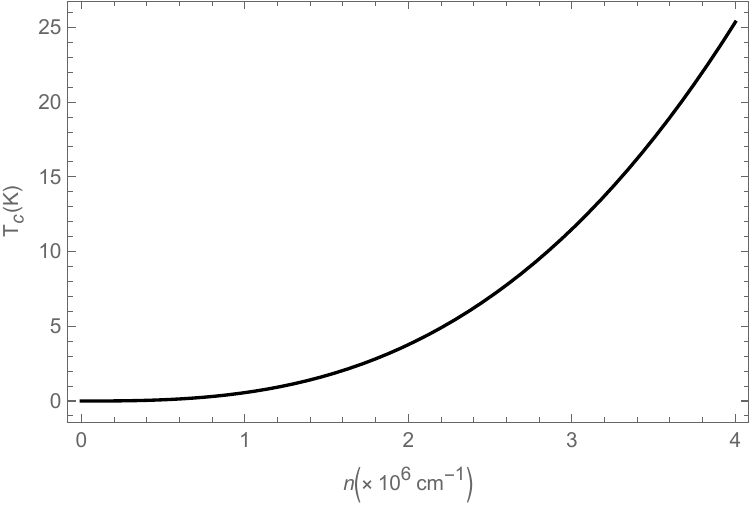}
\caption{}
\end{subfigure}
\begin{subfigure}[b]{0.40\textwidth}
\includegraphics[width=\textwidth]{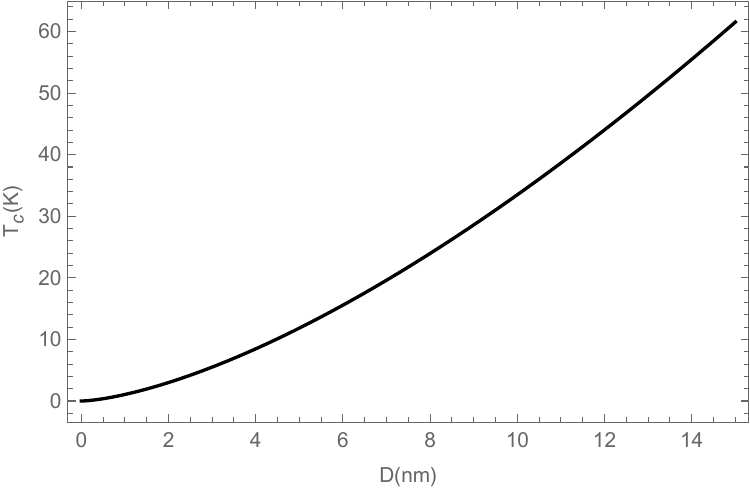}
\caption{}
\end{subfigure}
\end{center}
\caption{Critical temperature $T_c$ as a function of (a) the exciton density $n$, calculated for an inter-layer separation of $N_L = 7$ layers of $h$-BN ($D\approx 2.33 nm$); and (b) of the inter-layer separation $D$ for a system with concentration $n = 2 \times 10^6$ cm$^{-1}$. }\label{crit_temp}
\end{figure}

{ In Fig. \ref{crit_temp}, we present plots depicting the way in which the critical temperature $T_c$ depends on both the one-dimensional concentration of excitons $n$ (Fig. \ref{crit_temp}(a)) and on the layer separation $D$ (Fig.\  \ref{crit_temp}(b)). We note that our results predict the existance of a superfluid phase at relatively high temperatures. The value of $T_c$ increases monotonically with both the exciton concentration $n$, and the inter-layer separation $D$. }

\medskip
\par 
{In general, the exciton concentrations on the right and left edges of the double-layer strip are not the same. This means that, if $n_L \neq n_R$, we have two phase transitions at two distinct temperatures. If, for example, $n_R>n_L$, there is an interval of temperatures in which there is no superfluid phase in the left edge, but there is one in the right edge. By controlling $n_L$ and $n_R$, one can control on which edge of the strip superfluidity exists.}

{
\section{Experimental realization}\label{experimental}

\begin{figure} [h]
\includegraphics[width = 0.8\textwidth]{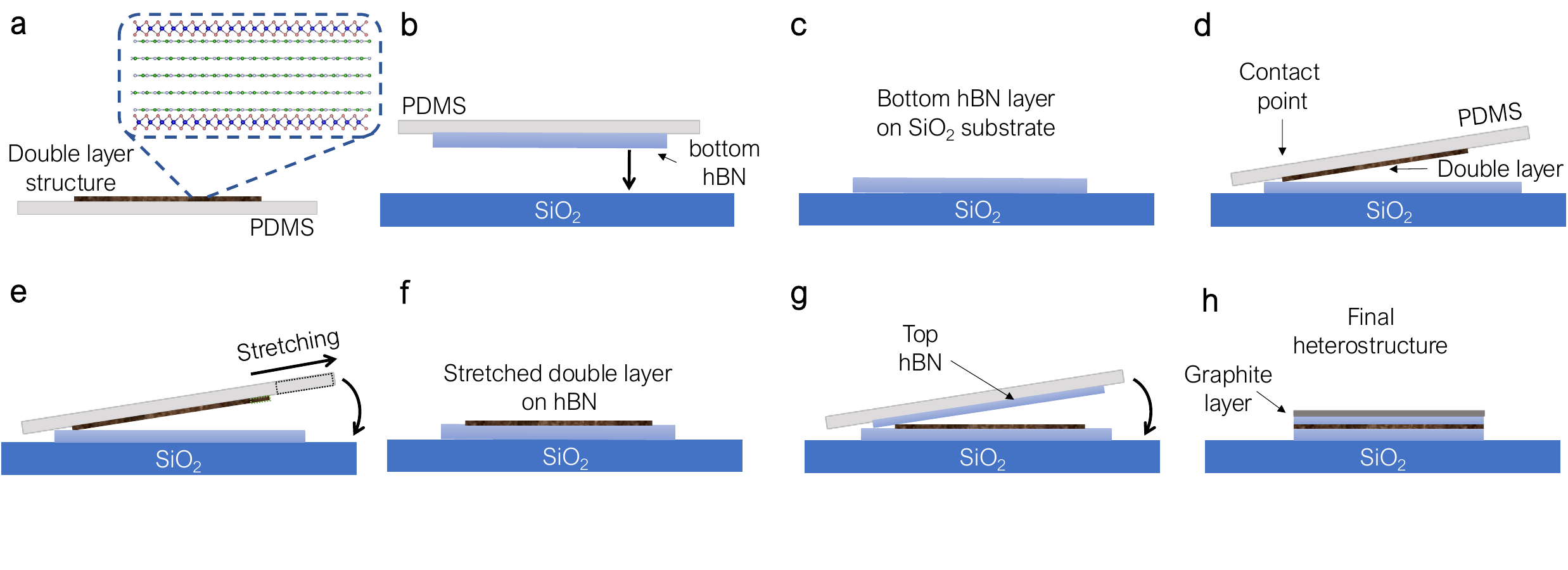}
\caption{Steps of the proposed fabrication method to experimentally observe dark exciton
superfluidity in TMDC double layer.}\label{exper}
\end{figure}

Experimental realization of superfluid states of dipolar dark excitons requires the creation of non-homogeneous strain landscapes in one-dimensional  TMDCs. Engineering a system like the one depicted in Fig. \ref{schem} (b) presents challenges, as it necessitates fabricating 1D TMDC strips and generating a strain gradient with radial symmetry. However, experimental observation of dark exciton superfluidity may be achieved in simpler structures featuring non-homogeneous strain with linear symmetry. In such a system, the vector potential varies linearly with spatial coordinates, resulting in a finite pseudomagnetic field.

\medskip
\par

Linear tensile strain gradients have so far been demonstrated, for example, in monolayers of WS$_2$ encapsulated with thin \textit{h}-BN layers \cite{grosso2}. Such strain-engineered systems can be obtained through mechanical exfoliation of TMDC layers using the standard scotch tape technique on thin polymer foils like polydimethylsiloxane (PDMS). Encapsulation with top and bottom \textit{h}-BN layers enhances exciton properties by stabilizing the electromagnetic and dielectric environment. A feasible fabrication pathway for a 2D heterostructure enabling the observation of dark exciton superfluidity is as follows.

\medskip
\par

First, a double layer stack consisting of two monolayers of TMDC separated by an \textit{h}-BN spacer is prepared using standard mechanical exfoliation and stacking techniques \cite{expsec1} (Fig. \ref{exper} (a)). The double layer structure is initially placed on a PDMS foil to facilitate its integration into the final heterostructure. The bottom \textit{h}-BN layer is then transferred onto a substrate, such as SiO$_2$/Si, serving as the base for the double layer (Fig. \ref{exper} (b,c)). Next, the double layer structure on PDMS is engaged on the bottom \textit{h}-BN layer at a large angle (Fig. \ref{exper} (d)). Upon contact, the PDMS foil and double layer are mechanically pulled to induce tensile strain before deposition onto the \textit{h}-BN (Fig. \ref{exper} (e)). Due to the mechanical stress in the PDMS, the double layer is transferred onto the substrate with a progressively increasing tensile strain (Fig. \ref{exper} (f)). To maintain the induced strain, a thin top \textit{h}-BN layer is then transferred onto the double layer stack at a large angle (Fig. \ref{exper} (g)). A thin graphite layer can subsequently be deposited on top to serve as the top electrode, enabling the application of a vertical electric field across the double layer (Fig. \ref{exper} (h)). As a 1D structure is required for exciton confinement, the final step involves patterning the heterostructure using standard nanolithography techniques, achieving resolutions down to a few nanometers. After defining the 1D strip of the strained double layer, metal electrodes are evaporated onto the structure to gate the top graphite layer and the bottom Si substrate. The resulting heterostructure will incorporate all key theoretical requirements, including a non-homogeneous tensile strain landscape in a double layer system, a 1D geometry, and the ability to apply an externally controlled vertical electric field.

\medskip
\par
The heterostructure illustrated in Fig. \ref{exper} (h) could serve as a suitable test bench for observing the superfluidity of dark excitons. Such a bosonic regime could be investigated at cryogenic temperatures through power-dependent measurements to evaluate the onset of macroscopic coherence in the dark exciton gas as a function of exciton density. Coherence could be further tested via interferometric measurements and the reconstruction of the first-order coherence function \cite{expsec2}.

}


\section{Discussion and Conclusion} 
\label{conc}

In summary, we have investigated the energy dispersion, binding energy, collective excitations, and superfluidity of excitons in {strained strips of double-layer TMDCs}. We considered nano heterostructures of two sheets of TMDC under an arc-shaped strain separated by a variable number of layers of \textit{h}-BN, as shown in Fig.\  \ref{schem}. { We have demonstrated  that provided the strength of strain is large enough, the presence of the strain field will create dark excitons in the system. Dark excitons appear whenever the maximum of the valence band and minimum of the conduction band for electrons correspond to different values of the wave-vector $q$. We have shown that the distance in $k$-space between these two values increases with strain.}   We have numerically calculated the energy spectrum of the excitons for the case of strained MoSe$_2$. We have seen that the binding energy of the excitons decreases mildly with the intensity of the strain. The momentum gap between the electron and hole, however, increases with strain, given that the radius of curvature is smaller than a cut-off value $R<R_c$, as can be seen in Fig. \ref{edispcondval}(b). Following this, we have obtained the energy spectrum for collective excitations. We have shown that the dipolar dark excitons formed in the system are characterized by a sound-like spectrum and we have obtained an  expression for the speed $c_S$ of sound for this system. Lastly, we have shown that this system will exhibit a superfluid phase and we calculated the mean-field critical temperature for superfluidity. {Additionally, we have shown that a superfluid flow and counterflow appear in the double-layer strip,  on opposite  edges. The two separate flows are a direct consequence of the excitons having a doubly degenerate ground state, with opposite momenta, each of which is located on one of the edges.}

\medskip
\par

In the present work, we treated the collective excitations in the mean-field approach. One could expect a phase transition between the superfluid phase and normal phase to be of the Kosterlitz-Thouless kind \cite{KT1,KT2}, and, therefore, to occur at a smaller temperature $T_c$, when the superfluid component of the system has a non-zero value. We intend to address such an issue in future work.
\medskip
\par

Experimental evidence for the existance of dark-excitons in TMDCs has been the subject of many recent studies \cite{darkexc1,darkexc2,darkexc3,darkexc4}. { In Ref. \cite{darkexc5}, it has been shown that {the lifetime of} dark excitons in single-layer WSe$_2$ is about 50 times longer than that for  bright excitons. In Ref. \cite{darkexc4} it has been shown that dark excitons in MoSe$_2$/WSe$_2$ heterostructures can have lifetimes of up to 2 microseconds.} {With the experimental techniques proposed here, one could achieve long-living superfluidity of strain-induced dark excitons in effectively one-dimensional strips.} 

\medskip
\par

\section*{Acknowledgements}

Oleg Berman acknowledges the support from the PSC CUNY Award No. 66382-00 54;
 Godfrey Gumbs and Gabriel Martins gratefully acknowledges funding from the U.S. National Aeronautics and Space Administration (NASA) via the NASA-Hunter College Center for Advanced Energy Storage for Space  under cooperative agreement 80NSSC24M0177. Gabriele Grosso acknowledges support from the National Science Foundation (NSF) (grant No. DMR-2044281).

\end{document}